# Dynamics of the multi-soliton waves in the sine-Gordon model with two identical point impurities

A. M. Gumerov[1*], R. V. Kudryavtsev[1], E. G. Ekomasov[1,2]

[1] Bashkir State University, Zaki Validi St., 32, Ufa, 450076, Russia

[2] National Research South Ural State University, Lenina prospect 76, Chelyabinsk 454080, Russia

e-mail: bgu@bk.ru*, xc.89@mail.ru, ekomasoveg@gmail.com.

**Abstract**

The particular type of four-kink multi-solitons (or quadrons) adiabatic dynamics of the sine-Gordon equation in a model with two identical point attracting impurities has been studied. This model can be used for describing magnetization localized waves in multilayer ferromagnet. The quadrons structure and properties has been numerically investigated. The cases of both large and small distances between impurities has been viewed. The dependence of the localized in impurity region nonlinear high-amplitude waves frequencies on the distance between the impurities has been found. For an analytical description of two bound localized on impurities nonlinear waves dynamics, using perturbation theory, the system of differential equations for harmonic oscillators with elastic link has been found. The analytical model qualitatively describes the results of the sine-Gordon equation numerical simulation.

**PACS numbers:** 02.30.Ik, 02.60.Cb, 05.45.Yv, 75.30.Hx

**Keywords:** nonlinear waves, solitons, multi-soliton, quadrons, sine-Gordon equation, impurity, perturbations, numerics, numerical simulation.

## 1. Introduction

Although solitons initially appeared in the integrable systems study, very soon they began to be used for non-integrable systems, describing many physical applications in hydrodynamics, condensed matter physics, field theory, etc. (see [1-3]). For example, the Sine-Gordon Equation (SGE) solitons describe the domain walls in magnetics, dislocations in crystals, fluxons in Josephson junctions and crossings, etc. (see [4].). Currently, the sine-Gordon equation and its three simple solutions: kink, breather and phonons, are well studied [1-4]. The researchers strong interest is attracted to the effect of disturbances, arising from the physical applications consideration, on the SGE soliton solutions structure, properties and generation [5-15]. If the study of the small perturbations influence on the SGE solutions can be carried out by a well-developed perturbation theory for solitons [2,7-8], then the influence of large perturbations in the general case can be carried out only with the help of numerical methods (see e.g. [8-10]). The researchers interest is attracted to the question of various types of disturbances influence on the soliton dynamics. Many papers are devoted to the study of the spatial modulation (heterogeneity) of the periodic potential (or presence of impurities in the system) influence on the SGE solitons dynamics. (see e.g. [1-2], [7-19]). The SGE model with impurities describes, for example, the case of a multilayer ferromagnet [20-22]. The importance of impurity modes for kink dynamics is shown in [1], [12-16]. The structure and properties of localized nonlinear waves excited on an impurity were analyzed numerically in [17]. The case of several deltashaped point impurities, which are of interest in some physical applications [18], and even the case of a spatially modulated harmonic potential [19] were considered. The possibility of exciting an impurity mode as a result of kink scattering was also taken into account, which leads to a considerable change in the kink dynamics [8-10].

Much less is known about the multi-soliton solutions of the SGE than is known about the soliton, and that is why multi-soliton are the topic of study in this paper. Earlier [23] there was found an interesting three-kinks SGE solution of wobble type, in [24] it was shown that such SGE solution can be excited at kink pinning in a single-attracting impurity. In the case of two identical impurities (or the case of five-layer ferromagnet) [21, 25], strong collective effects in the system were revealed, which can be used in kink pinning by the impurity to excite multi-solitons of the SGE. Additionally, it was shown that there can occur another interesting effect, namely, quasi-tunneling, in which case a kink passing through a double impurity needs less kinetic energy than that for passing a single impurity ofthe same sizes. However, the excitation, structure, and characteristics of SGE multi-solitons in the case of several impurities have not been completely investigated. This paper deals with the dynamics of exciting localized nonlinear waves of the multi-soliton type of the one-dimensional sine-Gordon model with two identical "point impurity". First, with the help of perturbation theory, there was found a system of differential equations for harmonic oscillators with elastic link, which describes the dynamics of two bound, localized on impurities, nonlinear waves. Next SGE multi-solitons with two impurities were numerically investigated. Then the cases of both large and small distances between impurities were viewed. The dependence of the localized in impurity region nonlinear high-amplitude waves frequencies on the distance between the impurities was found. Finally, the analytical model qualitatively describing the results of the sine-Gordon equation numerical simulation was found.

## 2. Main equations and analytical results

Let's consider a system defined by the Lagrangian:

$$L = \int_{-\infty}^{+\infty} \left\{ \frac{1}{2} u_t^2 - \frac{1}{2} u_x^2 - [1 - \varepsilon\delta(x) - \varepsilon\delta(x-d)](1 - \cos u) \right\} dx \qquad (1)$$

where the summand $\varepsilon\delta(x)$ simulates a point impurity, $\delta(x)$ — Dirac delta function, $\varepsilon$ — constant. The Lagrangian (1) corresponds to the motion equation for a scalar field $u(x,t)$ in the form:

$$u_{tt} - u_{xx} + \sin u = [\varepsilon\delta(x) + \varepsilon\delta(x-d)]\sin u \qquad (2)$$

Equation (2) – is the modified sine-Gordon equation (MSGE). The perturbation terms in the right side of equation (2) describe, for example, the case of five-layer ferromagnet with different values of the magnetic anisotropy in various layers [21].

Let's consider the structure and dynamics of localized nonlinear waves excited, for example, as a result of the kink scattering on point impurities. Equation (2) with zero right-hand side has a spatially localized solution — resting breather [2,4]

$$u(x,\tau) = 4\,\text{arctg}\left[\frac{\sqrt{1-\Omega_{single}^2}}{\Omega_{single}} \frac{\sin(\Omega_{single}\tau)}{\text{ch}((x-x_0)\sqrt{1-\Omega_{single}^2})}\right] \qquad (3)$$

where $\Omega_{single}$ — breather frequency, and $x_0$ — the coordinate of its center.

First, for the theoretical analysis of the structure and dynamics of the equation (2) localized solutions is applicable an approximate collective coordinate approach, previously used to analyze the impurity mode oscillation at a single point impurity [1], [2], is applicable. The presence of localized waves in the impurities area (or impurity modes) is taken into account through the introduction of two collective variables — $a_1 = a_1(t)$ and $a_2 = a_2(t)$ — which are the amplitudes of these waves. The expression for the impurity modes will be taken in the form similar to that used previously for the case of a single impurity [1,8]:

$$\begin{cases} u_1 = a_1(t)\exp(-\varepsilon|x|/2) \\ u_2 = a_2(t)\exp(-\varepsilon|x-d|/2) \end{cases} \qquad (4)$$

In the small oscillations approximation considering that $a_i(t) = a_{i0}\cos(\Omega_{single}t + \theta_0)$, where $\theta_0$ — initial phase. When solving (2) for the case of a single impurity the following expression may be obtained for the impurity mode frequency — $\Omega_{single} = (1 - \varepsilon^2/4)^{1/2}$. The general solution of the problem — $u_{ansatz}$ — will be searched in the form:

$$u_{ansatz} = u_1 + u_2 = a_1(t)\exp(-\varepsilon|x|/2) + a_2(t)\exp(-\varepsilon|x-d|/2) \qquad (5)$$

To find an analytical solution we assume that $a_1(t)$ and $a_2(t)$ are small enough (are order of magnitude $\varepsilon$). Then we can assume within the approximation that:

$$u_{ansatz} \ll 1 \qquad (6)$$

and the nonlinear term in the Lagrangian (1) can be expanded in Taylor series to the second-order terms in $\varepsilon$ [2]:

$$\cos u - 1 \approx -\frac{1}{2}u^2 \qquad (7)$$

Substituting (5) into (1) based on the approximation (7) leads after integration to a new effective Lagrangian already dependent on new variables $a_1$ and $a_2$:

$$L \approx \frac{\dot{a}_1^2(t)}{\varepsilon} + \frac{\dot{a}_2^2(t)}{\varepsilon} + \dot{a}_1(t)\dot{a}_2(t)\frac{(2+\varepsilon d)e^{-\frac{\varepsilon d}{2}}}{\varepsilon} + a_1^2(t)\left[\frac{\varepsilon}{4} - \frac{1}{\varepsilon} + \frac{\varepsilon}{2}e^{-\varepsilon d}\right]$$
$$+ a_2^2(t)\left[\frac{\varepsilon}{4} - \frac{1}{\varepsilon} + \frac{\varepsilon}{2}e^{-\varepsilon d}\right] + a_1(t)a_2(t)\left[\frac{3\varepsilon}{2} + \frac{\varepsilon^2 d}{4} - \frac{2}{\varepsilon} - d\right]e^{-\frac{\varepsilon d}{2}} \quad (8)$$

The equations of motion for $a_1(t)$ and $a_2(t)$ can be obtained by inserting the obtained effective Lagrangian into the Lagrange equations system of the second kind:

$$[\ddot{a}_1(t)]\,[4 - (2+\varepsilon d)^2 e^{-\varepsilon d}]/\varepsilon$$
$$+ a_1(t)\left[\frac{4}{\varepsilon} - \varepsilon + \left(\varepsilon + 2\varepsilon^2 d - \frac{4}{\varepsilon} - 4d + \frac{\varepsilon^3 d^2}{4} - \varepsilon d^2\right)e^{-\varepsilon d}\right]$$
$$- a_2(t)\varepsilon[2 - (2+\varepsilon d)e^{-\varepsilon d}]e^{-\frac{\varepsilon d}{2}} = 0 \quad (9a)$$

$$[\ddot{a}_2(t)]\,[4 - (2+\varepsilon d)^2 e^{-\varepsilon d}]/\varepsilon$$
$$+ a_2(t)\left[\frac{4}{\varepsilon} - \varepsilon + \left(\varepsilon + 2\varepsilon^2 d - \frac{4}{\varepsilon} - 4d + \frac{\varepsilon^3 d^2}{4} - \varepsilon d^2\right)e^{-\varepsilon d}\right]$$
$$- a_1(t)\varepsilon[2 - (2+\varepsilon d)e^{-\varepsilon d}]e^{-\frac{\varepsilon d}{2}} = 0 \quad (9b)$$

The equations system (9) for $a_1(t)$ and $a_2(t)$ is lengthy. Let's introduce the following notation:

$$F = F(\varepsilon, d) = \frac{\varepsilon^2(2 - (2+\varepsilon d)e^{-\varepsilon d})e^{-\frac{\varepsilon d}{2}}}{4 - (2+\varepsilon d)^2 e^{-\varepsilon d}} \quad (10)$$

$$\Omega_0^2 = \Omega_0^2(\varepsilon, d) = 1 - \frac{\varepsilon^2}{4} - \frac{\varepsilon^2\left(1 + e^{-\frac{\varepsilon d}{2}}\right)e^{-\frac{\varepsilon d}{2}}}{2 + (2+\varepsilon d)e^{-\frac{\varepsilon d}{2}}} \quad (11)$$

Using (10) and (11), multiplying both equations of the equations system (9) by $\varepsilon[4 - (2+\varepsilon d)^2 \exp(-\varepsilon d)]^{-1}$ we obtain:

$$\begin{cases} [\ddot{a}_1(t)] + a_1(t)\Omega_0^2 = F(a_2(t) - a_1(t)) \\ [\ddot{a}_2(t)] + a_2(t)\Omega_0^2 = -F(a_2(t) - a_1(t)) \end{cases} \quad (12)$$

The equations system (12) is a system of ordinary differential equations of the second order. It follows from (12) that system oscillations of the breather type bound localized waves in the small oscillations approximation can be described by a system of coupled effective harmonic oscillators with the same proper frequency $\Omega_0(\varepsilon,d)$. Each of them undergoes the external force $(a_2(t) - a_1(t))F(\varepsilon,d)$ (of elastic type) from the "neighboring" oscillator. In this case, the coupling coefficient $F(\varepsilon,d)$ can vary, for example, by changing parameter $d$ — the distance between impurities. Let's analyze the behavior of functions (10), (11) in the limiting case $d \to \infty$:

$$\lim_{d \to \infty} F(\varepsilon, d) = 0 \quad (13)$$

$$\lim_{d \to \infty} \Omega_0^2(\varepsilon, d) = 1 - \frac{\varepsilon^2}{4} = \Omega_{single}^2 \quad (14)$$

Expression (13) shows that with the parameter increase, the coupling coefficient is reduced to zero, and (12) goes into the equations system for disconnected harmonic oscillators. From (14) one can see that $\Omega_0$ goes into the expression for a single impurity. In this case, the oscillators oscillate independently at the proper frequency of a single impurity.

Let's study the dependence of the expressions (10) and (11) on the parameter $d$. Fig. 1 shows that at $d \to \infty$ both functions $\Omega_0$ and $F$ asymptotically tend to the corresponding values obtained in (13) and (14). In case $d \to 0$ this function begins to strongly increase, which corresponds to the increase of the link "rigidity" between the effective oscillators. Meanwhile value $\Omega_0(\varepsilon, d = 0)$ has the final value corresponding to the "dublicated" effective oscillator, i.e. in the expression for $\Omega_{single}$ one has to insert value $2\varepsilon$.

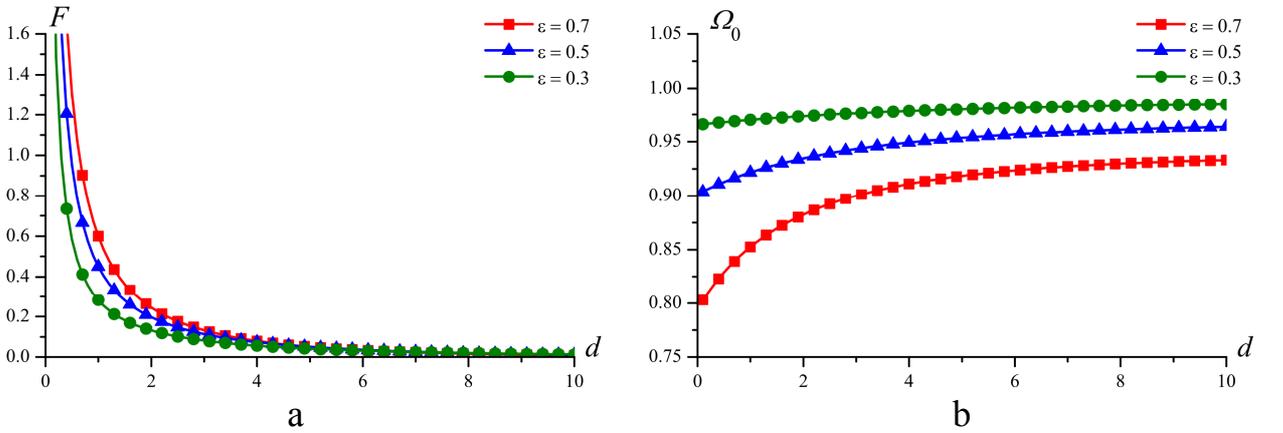

Fig. 1. The dependence of system (12) parameters on the distance between impurities $d$: a) $F(\varepsilon,d)$; b) $\Omega_0(\varepsilon,d)$.

As is known, the system of the harmonic oscillators linked by elastic coupling may have solutions in the form of in-phase or antiphase oscillations. To analyze these solutions properties let us move to the new variables:

$$\varphi_S = \frac{a_1(t) + a_2(t)}{\sqrt{2}} \qquad \varphi_A = \frac{a_1(t) - a_2(t)}{\sqrt{2}} \qquad (15)$$

By adding and subtracting equations in the system (12) using (15) one can come to the following system of equations:

$$\begin{cases} \ddot{\varphi}_S + \Omega_S^2 \varphi_S = 0 \\ \ddot{\varphi}_A + \Omega_A^2 \varphi_A = 0 \end{cases} \qquad (16)$$

where $\Omega_S = \Omega_0$, $\Omega_A = (\Omega_0 + 2F)^{1/2}$. Thus, the oscillators fluctuations (12) can be regarded as a superposition of oscillations with symmetric $\Omega_S$ and anti-symmetric modes $\Omega_A$, which are commonly referred to as normal frequencies (modes) of the system and introduced variables (15) — normal coordinates. In special cases, when the oscillations are in phase or antiphase, the whole system, and each oscillator in particular, oscillate in the corresponding normal frequency. For clarity, let us consider a few special cases with specially selected initial conditions. For example, at time zero both effective oscillators deflect from their equilibrium position on $a_{0S}$ and have the same speed $v_{0S}$. In this case, the movements will be symmetrical (effective spring is not stretched) and are given in the following expression:

$$a_1(t) = a_2(t) = a_{0s} \cos \Omega_S t + \frac{\dot{a}_{0s}}{\Omega_S} \sin \Omega_S t \tag{17}$$

If at time zero both effective oscillators are deviated by $a_{0A}$ in different directions and have the same speed $v_{0A}$ antisymmetric vibrations are excited:

$$a_1(t) = -a_2(t) = a_{0a} \cos \Omega_A t + \frac{\dot{a}_{0a}}{\Omega_A} \sin \Omega_A t \tag{18}$$

At an arbitrary choice of the initial conditions, the system motion is described by a superposition (17) and (18):

$$\begin{cases} a_1(t) = a_{0s} \cos \Omega_S t + \dfrac{\dot{a}_{0s}}{\Omega_S} \sin \Omega_S t + a_{0a} \cos \Omega_A t + \dfrac{\dot{a}_{0a}}{\Omega_A} \sin \Omega_A t \\ a_2(t) = a_{0s} \cos \Omega_S t + \dfrac{\dot{a}_{0s}}{\Omega_S} \sin \Omega_S t - a_{0a} \cos \Omega_A t - \dfrac{\dot{a}_{0a}}{\Omega_A} \sin \Omega_A t \end{cases} \tag{19}$$

Solutions (19) can be obtained by solving the equations system (16) while returning to the original variables (15).

Let's consider possible fluctuations in system (12) by means of numerical integration by the fourth-order accuracy Runge-Kutta method at different initial conditions. Let us analyze the oscillations nature depending on parameter $d$ and initial conditions. This will allow to test the correctness and accuracy of the frequency characteristics calculation using the numerical method, and to identify some of the features that occur at frequencies determination. In this case, we assume for simplicity $\dot{a}_1(t = 0) = 0$; $\dot{a}_2(t = 0) = 0$, and will change in calculation only $a_1(t = 0) = a_{01}$; $a_2(t = 0) = a_{02}$. Next let's consider three cases corresponding to different initial conditions. We use Fourier-decomposition into frequency components to define $\Omega_S$ and $\Omega_A$.

The results of dependence numerical simulation and Fourier-analysis are shown in Fig. 2. The initial conditions are chosen so that the case in Fig. 2a is close to the phase oscillations, so the amplitude of the first frequency component in the Fourier spectrum is significantly higher than the second. Thus we can say that in a movements superposition (19) symmetrical summands are dominating. Another situation is observed in Fig. 2c, where anti-symmetric summands are dominating. Thus, in particular cases, when the fluctuations nature is too close to either symmetric or antisymmetric, there can appear problems with one of the frequency components definition. Therefore, the best, in terms of analyzing frequency components, are the cases similar to that shown in Fig. 2b. In addition the calculated frequencies show good quantitative agreement with analytical expressions $\Omega_S$ and $\Omega_A$.

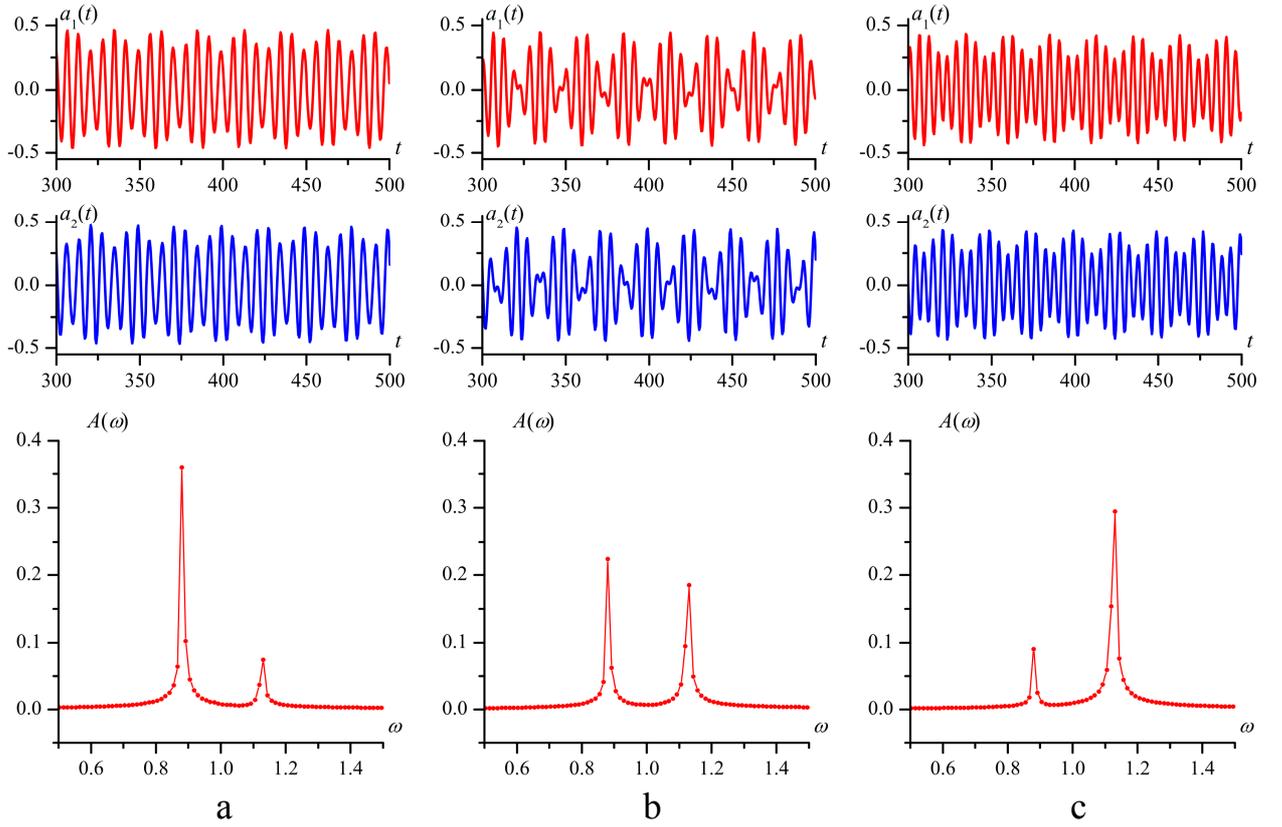

Fig. 2. The dependence of the deviation value from the equilibrium position of the first $a_1(t)$ and second $a_2(t)$ oscillator on time $t$, resulting from numerical solution of system (12) and corresponding $a_1(t)$ discrete Fourier expansion $A(\omega)$. Parameters: $\varepsilon = 0.7$, $d = 2$. Initial conditions: a) $a_{01} = 0.5$, $a_{02} = 0.3$, b) $a_{01} = 0.5$, $a_{02} = 0$, c) $a_{01} = 0.5$, $a_{02} = -0.3$.

We next consider the effect of parameter $d$, that controls the stiffness of the connection between effective oscillators (see. Fig. 1a), on the emerging oscillation modes. For example, we present the results obtained only for the case $\varepsilon = 0.7$. The calculations showed that other considered cases are similar to the given one. Let's introduce in the initial conditions the initial phase difference of the oscillators (to excite oscillations in the beat mode). Fig. 3 shows three different cases corresponding to different values of the parameter $d = \{0.3, 1, 3\}$, which by the nature of frequency spectra $A(\omega)$ can be (conditionally) assigned to different oscillation modes.

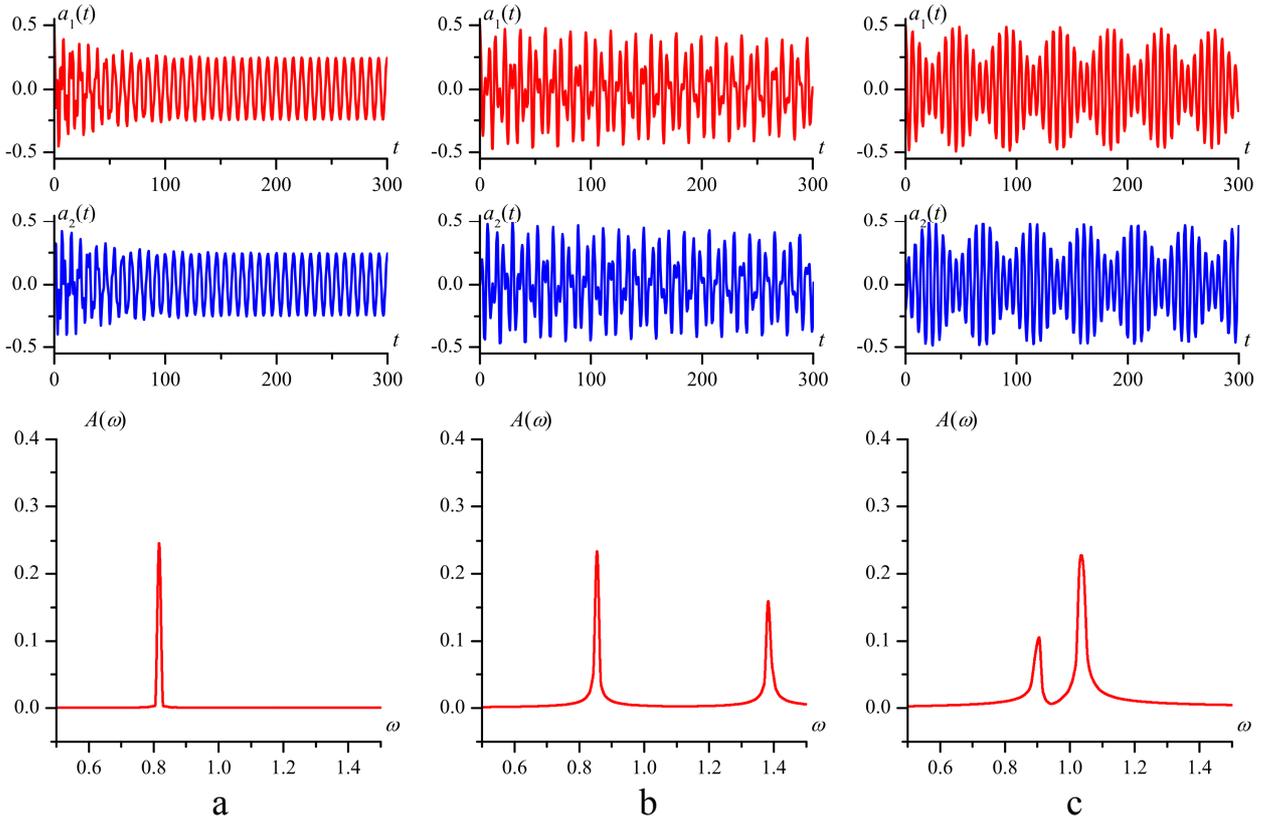

Fig. 3. The dependence of the deviation value from the equilibrium position of the first $a_1(t)$ and second $a_2(t)$ oscillator on time $t$, resulting from numerical solution of system (12) and corresponding $a_1(t)$ discrete Fourier expansion $A(\omega)$. Parameters: $\varepsilon = 0.7$, a) $d = 0.3$, $a_{01} = 0.5$, $a_{02} = 0$, b) $d = 1$, $a_{01} = 0.5$, $a_{02} = 0$, c) $d = 3$, $a_{01} = 0.5$, $a_{02} = -0.2$.

In the case of short-range distances between impurities $d = 0.3$ (Fig. 3a) it is clear that the link is very hard (strong). In any initial conditions of the oscillators vibrations, the oscillations phase difference is reduced to zero, and after a certain period of time oscillators begin to oscillate in phase (at a single frequency). The antisymmetric mode becomes not profitable and in reality this oscillation mode is not stable. In the case of small distances between impurities $d = 1$ (Fig. 3b) it is clear that the link is of medium hardness and there occur strong beats. Frequency components in spectrum $A(\omega)$ are located far from each other. From Fig. 3c for the case of large distance between impurities $d = 3$ one can see that the link is weak, and system beats are weak. Frequency components in spectrum $A(\omega)$ are located much closer. This oscillation mode corresponds to the "weak link" case (see. e.g. [26,27]), and general solution (19) can be simplified for it.

## 3. Numerical results

To check the domain of applicability of the analytical model and system of equations (12), obtained by perturbation theory for solitons (collective coordinate method), let's investigate numerical structure and dynamics of localized nonlinear waves from the original equation (2). To date, there developed quite a number of methods for the numerical solution of such nonlinear differential equations. For example, a number of studies [28-29] use spectral and pseudospectral Fourier methods for solving equations SGE. In paper [30] a compact finite-difference scheme and DIRKN-method are used. The method of lines is used in [31]. In this paper the method of finite differences for the numerical solution of the equation (2) is used. There has been selected a three-layer explicit solution scheme with derivatives approximation on a five-point pattern of the "cross" type. It was previously applied to simpler SGE modifications (see. for example [10], [17]). Let $\Delta x$ — be a coordinate step, $\tau$ — time step. This numerical scheme of the second approximation order at $\Delta x$ and $\tau$ has conditional stability when $(\tau/\Delta x) \leq 1/2$. In this case, the circuit is "one-step", uses a relatively small number of memory accesses and has the potential to optimize computational algorithm. In addition, the scheme used is convenient because it can be with minimal changes adapted both for other modifications of one-dimensional equation (2), and for multidimensional SGE variants.

Created algorithm for the numerical solution of the equation (2) works as follows. At the initial time, we have an SGE kink moving by inertia with a fixed velocity of the form

$$u_0(x,t) = 4\arctg\left[\exp\left(\pm\gamma(x - x(t))\right)\right], \qquad (20)$$

where $\gamma = (1 - v^2)^{-1/2}$, $v$ — kink moving velocity $x(t) = vt + x_0$ — kink center coordinate. The boundary conditions are: $u(-\infty, t) = 0$, $u(+\infty, t) = 2\pi$, $u'(\pm\infty, t) = 0$. Using the grid of coordinate from $N = 10^4$ knots, taking time as an iterative parameter and following the conditions of the explicit scheme convergence, the value of the function $u(x,t)$ in the next time was calculated. From the found function $u(x,t)$ we received the main characteristics of the nonlinear wave. Numerical experiment showed that after the kink passage there remained a nonlinear wave in each impurity, described by the oscillating function $u(x,t)$ of the bell type. For the case of one attracting impurity there has been shown previously [17] that the localized nonlinear wave can be regarded as a resting breather. The amplitude of the excited breather depends on the kink velocity value. The breather oscillation frequency — $\omega_{breather}$ practically does not depend on the kink velocity and is determined by the impurity parameters. As localized impurity waves related dynamics is observed, the resulting MSGE solution can be called a four-kinked multi-soliton.

Let's calculate frequency dependences of the localized waves oscillations depending on the distance between the impurities. Since localized waves are excited by the kink passage, so its initial velocity $v_0$ that determines phases initial difference between the localized waves. It is important to note that it is impossible to set optional oscillations phases initial difference by changing $v_0$. Consequently, it is impossible to excite the entire spectrum of possible states. For example, it is not always possible to excite anti-symmetric oscillation modes.

Fig. 4 solid lines plot analytically calculated symmetrical (curve 2) and antisymmetric (curve 1) mode frequencies of the system (12), and point lines — symmetric (curve 4) and antisymmetric (curve 3) fluctuations frequencies of the MSGE breathers. These frequency components have been calculated using the Fourier analysis method described above. Fig. 4 shows that there is qualitative agreement between the respective curves. The discrepancy between the curves for the parameter small values is primarily due to analytical expressions discrepancy for $\Omega_0(\varepsilon,d)$ and $F(\varepsilon,d)$ with the results of numerical calculation of the MSGE breathers oscillations (Fig. 5). That is, the expressions (10) and (11) describe well the link between the effective oscillators and their own frequency only at relatively large values of $d$. However, the curves, obtained numerically from MSGE, can be approximated in the viewed area by a simple exponential dependence (Fig. 5, curves 3.) of the form:

$$f(x) = A + Be^{Cx} \qquad (21)$$

Fig. 4 represents three qualitatively different parameter regions (similar to the cases considered in Fig. 3).

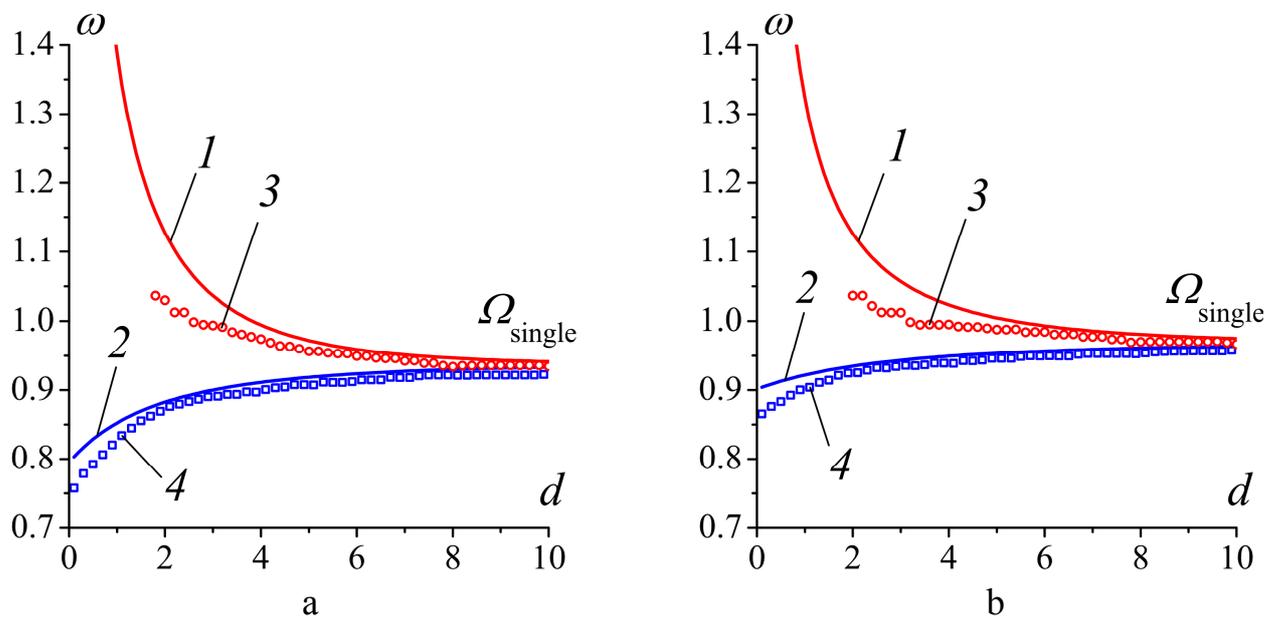

Fig. 4. Areas of possible values of the localized waves oscillations frequencies depending on distance between the impurities $d$ when: a) $\varepsilon = 0.5$ and b) $\varepsilon = 0.7$. 1 – in-phase oscillations frequency $\Omega_S$ ($a_{02} = 0.5$), 2 – antiphase oscillations frequency $\Omega_A$ ($a_{02} = -0.5$), 3 – MSGE localized waves in-phase oscillations frequency, 4 – MSGE localized waves antiphase oscillations frequency

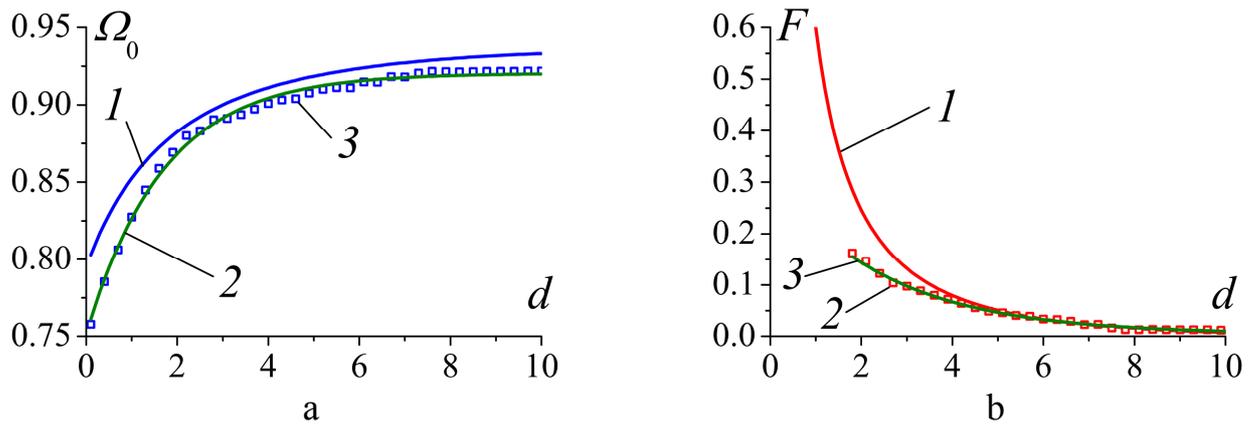

Fig. 5. The dependence of the parameters; a) $\Omega_0(\varepsilon,d)$; b) $F(\varepsilon,d)$ on the distance between the impurities $d$ at $\varepsilon = 0.7$. Curve 1 is calculated by analytic expressions a) (10) b) (11), points in curve 2 were obtained by numerical simulation of the MSGE localized waves oscillations, curve 3 – approximation of the numerically obtained points by expression (21): a) $A = 0.92048$, $B = -0.16906$, $C = -0.5878$; b) $A = 0.00417$, $B = 0.31079$, $C = -0.39969$.

In zone I ($d < 2$) only the in-phase mode was excited by kink passage. Obviously, the antiphase mode excitation requires much higher interaction energy. In zone II ($2 < d < 4$) there is a strong interaction between the localized impurity waves with periodic energy pumping. In zone III ($d > 4$) increase of $d$ results in localized impurity waves potential frequencies asymptotical aspiration to $\Omega_{single}$. In this zone, by changing $v_0$, one can excite both in-phase and antiphase oscillations. As a fundamental difference in regions II and III there can be distinguished the fact that parameter $v_0$ variation allows to initiate the states which are much closer to a symmetric mode.

## 4. Conclusions

Collective effect of two identical point impurities on the structure and dynamics of coupled nonlinear waves localized in the area of impurities has been studied. The case of exciting on both localized waves impurities (four-kinked multi-solitons) has been viewed. Analytically, for the case of small amplitudes, it was found that their oscillations can be described by the system of two harmonic oscillators bound by the elastic link. The structure and dynamics of localized waves at large and small distances between impurities has been investigated. The dependence of the frequencies, localized in the nonlinear high-amplitude waves impurity region, on the distance between the impurities has been found. It has been shown that the analytical model qualitatively describes the results of numerical simulation of the modified sine-Gordon equation for optional in size amplitudes of the localized waves. It has been shown that studied model can be used for describing magnetization localized waves in the multilayer ferromagnet.